# MARS spectral molecular imaging of lamb tissue: data collection and image analysis

R Aamir[a] *, A Chernoglazov[b], C J Bateman[a], A P H Butler[a,c,d,e], P H Butler[c,e,f], N G Anderson[a], S T Bell[e], R K Panta[a], J L Healy[g], J L Mohr[a], K Rajendran[a], M F Walsh[a], N de Ruiter[b], S P Gieseg[g], T Woodfield[h], P F Renaud[i], L Brooke[f], S Abdul-Majid[j], M Clyne[k], R Glendenning[i], P J Bones[d], M Billinghurst[b], C Bartneck[b], H Mandalika[b], R Grasset[b], N Schleich[m], N Scott[n], S J Nik[f], A Opie[a], T Janmale[g], D N Tang[f], D Kim[f], R M Doesburg[f,c], R Zainon[f], J P Ronaldson[a], N J Cook[o], D J Smithies[f], K Hodge[a]

[a] *Centre for Bioengineering, Department of Radiology, University of Otago Christchurch, NZ*
[b] *HIT lab NZ, University of Canterbury, Christchurch, NZ*
[c] *European Centre for Nuclear Research (CERN), Geneva, Switzerland*
[d] *Dept of Electrical & Computer Engineering, University of Canterbury, Christchurch, NZ*
[e] *MARS Bioimaging Ltd, Christchurch, NZ*
[f] *Dept of Physics & Astronomy, University of Canterbury, Christchurch, NZ*
[g] *Dept of Biochemistry, University of Canterbury, Christchurch, NZ*
[h] *Dept of Orthopaedic Surgery & MSM, University of Otago Christchurch, NZ*
[i] *Dept of Mathematics & Statistics, University of Canterbury, Christchurch, NZ*
[j] *Dept of Nuclear Engineering, King Abdulaziz University, Jeddah, KSA*
[k] *ILR, Christchurch, NZ*
[l] *Scott Technology Limited, Christchurch, NZ*
[m] *Dept of Radiation Therapy, University of Otago, Wellington, NZ*
[n] *Dept of Medicine, University of Otago, Christchurch, NZ*
[o] *Dept of Medical Physics and Bio-Engineering, CDHB, Christchurch, NZ*

 *E-mail:* `aamir.raja@otago.ac.nz`

ABSTRACT: Spectral molecular imaging is a new imaging technique able to discriminate and quantify different components of tissue simultaneously at high spatial and high energy resolution. Our MARS scanner is an x-ray based small animal CT system designed to be used in the diagnostic energy range (20 – 140 keV). In this paper, we demonstrate the use of the MARS scanner, equipped with the Medipix3RX spectroscopic photon-processing detector, to discriminate fat, calcium, and water in tissue. We present data collected from a sample of lamb meat including bone as an illustrative example of human tissue imaging. The data is analyzed using our 3D Algebraic Reconstruction Algorithm (MARS-ART) and by material decomposition based on a constrained linear least squares algorithm. The results presented here clearly show the quantification of lipid-like, water-like and bone-like components of tissue. However, it is also clear to us that better algorithms could extract more information of clinical interest from our data. Because we are one of the first to present data from multi-energy photon-processing small animal CT systems, we make the raw, partial and fully processed data available with the intention that others can analyze it using their familiar routines. The raw, partially processed and fully processed data of lamb tissue along with the phantom calibration data can be found at [http://hdl.handle.net/10092/8531].

---

\* Corresponding author.

**Contents**

## 1. Introduction

To perform spectral imaging, the Medipix All Resolution System (MARS) scanner uses a conventional CT broad spectrum x-ray source operated at a single accelerating voltage and a detector that can discriminate the energy of individual photons interacting with the detector sensor layer [1]. By combining this information with information about the energy dependence of x-ray attenuation in different materials, it is possible to discriminate different materials within the same voxel. The objective of this paper is twofold: first, to present unprocessed and processed data from our MARS scanner using the CdTe Medipix3RX camera in Charge Summing Mode (CSM) with four energy bins; and second, to demonstrate soft tissue imaging by the quantification of the fat content in tissue.

A substantial amount of work has been done on K-edge imaging of high Z contrast materials, demonstrating the ability of spectral technology to distinguish contrast materials from body tissues by their atomic spectra [2-9]. However, comparatively less research has been done to highlight the potential advantages of spectral imaging of tissue without using high Z contrast agents. Soft tissues such as fat and muscle (effective Z ≈ 7.5) are composed of low atomic number elements (e.g., hydrogen, Z = 1; carbon, Z = 6; nitrogen, Z = 7; and oxygen, Z = 8). At energies relevant to human diagnostic imaging (20 – 140 keV), x rays interact predominantly by a combination of the photoelectric (approximately proportional to $Z^3/E^3$ for E far from a K-edge) and Compton effects (approximately proportional to $1/E$) and therefore the probability of photoelectric absorption for Z < 10 at higher energies (> 50keV) is close to zero [10].

### 1.1 MARS Spectral Scanner

Our team in New Zealand has developed the world's first MARS spectral molecular scanner. It consists of a rotating gantry, MARS camera, cabinet controller, x-ray source, computer hardware and software as shown in Figure 1. The camera, with its software, is designed so it can be used in a stand-alone mode. The camera consists of three main components; 1) a bias voltage board, which provides a programmable bias voltage to the sensor layer from zero to 800V, 2) a chip carrier board on which the ASIC and sensor layer assembly is mounted, and 3) a readout board that allows connection between the ASIC and a host computer at a network speed of one gigabit. The current camera can accommodate up to six Medipix3 detectors. The scanner is designed to be highly modular for manufacturing, maintenance, service and upgrade reasons [1, 11, 12]. Scans are accomplished by continuous gantry rotation and requires less than 1 min x-ray exposure for 360 acquisitions in one complete rotation using a 40ms exposure time.

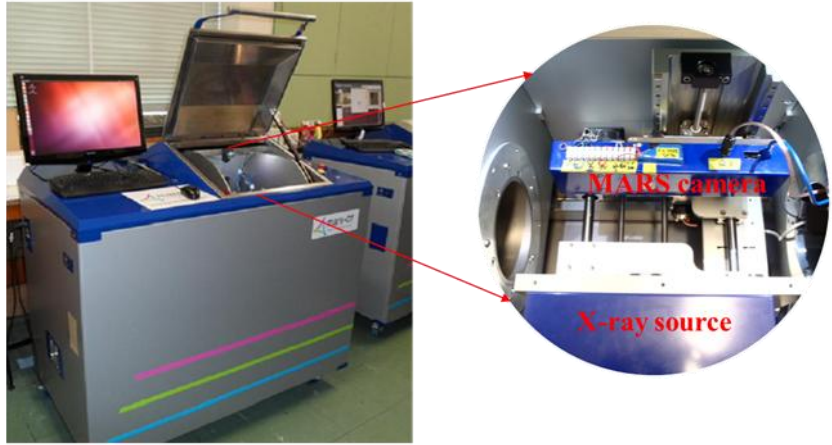

**Figure 1** *Left:* **A photo of one of MARS spectral molecular scanners.** *Right:* **Inside view of the MARS gantry showing the MARS camera and x-ray source. The sample is placed between the camera and x-ray source in the scanner such that its projection falls on the camera.**

The Medipix "flip-chip" design permits various sensors to be bonded to the Medipix ASIC [13]. Sensor materials include Silicon (Si), Gallium Arsenide (GaAs), Cadmium Telluride (CdTe) and Cadmium-Zinc Telluride (CZT); each with different attenuation characteristics for preclinical and clinical x-ray energy ranges. Silicon is traditionally used in preclinical setups due to the availability of good quality homogeneous material, and high charge carrier transport properties (1 $cm^2$/V). However, 300 μm Si sensors offer detection efficiency less than 30% for energies above 20keV and less than 5% above 40 keV due to its low atomic number (Z = 14). High energy x-ray photons require high Z sensors such as CdTe (Z = 48, 52) to provide an adequate absorption [14].

Despite improvements in the quality of high Z crystal materials and in readout electronics, photon counting detectors are susceptible to charge sharing [15-17] due to small pixel pitches (if less than about 300μm-side) making them technically challenging [18, 19]. Charge sharing was addressed for the first time in the Medipix3 chip [20]. However, due to a design flaw in early variants prevented the use of this new feature [21, 22]. The Medipix3RX ASIC has been designed to eliminate these flaws to provide accurate energy information [23, 24]. The Charge Summing feature included in the Medipix3RX ASIC allows the total charge from a single interaction to be summed and allocated to the closest pixel. To do this the charge from the interaction is summed across clusters of four pixels and each pixel communicates with its neighbours to locate the pixel with the highest charge for a co-incident event, the total charge is then allocated to this pixel [25].

Another feature of Medipix3RX is spectroscopic imaging. This has been achieved by introducing a second counter channel into each pixel enabling two energies per pixel. In addition, clusters of four pixels can communicate to act as a single larger pixel with eight low thresholds. This spectroscopic mode is a second key requirement for spectral molecular imaging.

**1.2 Spectral Molecular Imaging**

Spectral CT is specific, non-invasive, and quantitative. As each material has a specific measurable x-ray spectrum, spectroscopic imaging can simultaneously measure several biomarkers of biological processes at the cellular and molecular level, using simultaneously acquired data for multiple energy bins [26]. The combination of high spatial and spectral resolution with specific identification and quantification of multiple tissue components, non-invasively, is unique to specific cells and molecules. This cellular and molecular specific imaging is known as *Spectral Molecular Imaging* [27]. It is the quantification which is the key component of molecular imaging. It is hard to predict the full clinical significance of spectral molecular imaging, but an overview can be gained by looking into the pre-clinical results of applications such as, spectral imaging of atherosclerotic plaque [28, 29], soft tissue quantification [30-33], imaging with functionalized gold nanoparticles to assess plaque vulnerability [26] and various other research areas [34, 35].

This work is published with the intention of allowing interested parties access to raw, partially processed and fully processed data of a lamb tissue sample produced using the Medipix3RX camera within the MARS scanner platform. Full access to the data will allow users to test and compare their own processing routines. The data is analysed using the MARS-ART 3D reconstruction routine [36] and a post-reconstruction constrained linear least squares material decomposition algorithm.

**2. Experimental Setup**

In this experiment, a meat specimen was prepared from a fresh lamb chop, which included muscle (water-like), fat (lipid-like) and bone (calcium-like) regions, and scanned as shown in Figure 2. We used a 2 mm thick CdTe sensor (128×128), bump-bonded at 110 μm to Medipix3RX ASIC provided to us by X-ray Imaging Europe GmbH (http://www.xi-europe.com) and installed in the MARS-CT4 scanner. In charge summing mode (CSM), 720 circular projections over 360° were acquired using a Source-Ray SB-80-1K x-ray tube (Source-Ray Inc, Ronkonkoma, NY) with a tungsten anode having 1.8-mm-Al equivalent intrinsic filtration. The focal spot size was ~33 μm. Several vertical positions of the CdTe Medipix3RX camera were used to create a virtual detector to cover the 23 mm field of view (FOV). The bias voltage applied to the sensor was −440V. The source to detector distance (SDD) was 131.8mm and object to detector distance (ODD) was 48mm. The magnification factor of ~1.1 has been used for this experiment. Camera readout was performed using the MARS readout system [22].

Before the measurements, threshold equalization with respect to the noise edge, and energy calibration of the detector, were performed. Flat-field measurements (500 flat-fields per energy bin) were taken before specimen scanning to correct for variations in pixel response. Dark-field images (50 dark-fields per energy bin) were also acquired before and after the scan. The tube was operated at 50 kVp with a current of 120 μA and using four low energy thresholds (15, 20, 25, 30 keV). The exposure time of each acquisition was 40 ms. For data analysis and HU calibration, a phantom having $CaCl_2$ (320 mg/ml) and lipid (vegetable oil), along with air and water, was also scanned with the same parameters mentioned earlier.

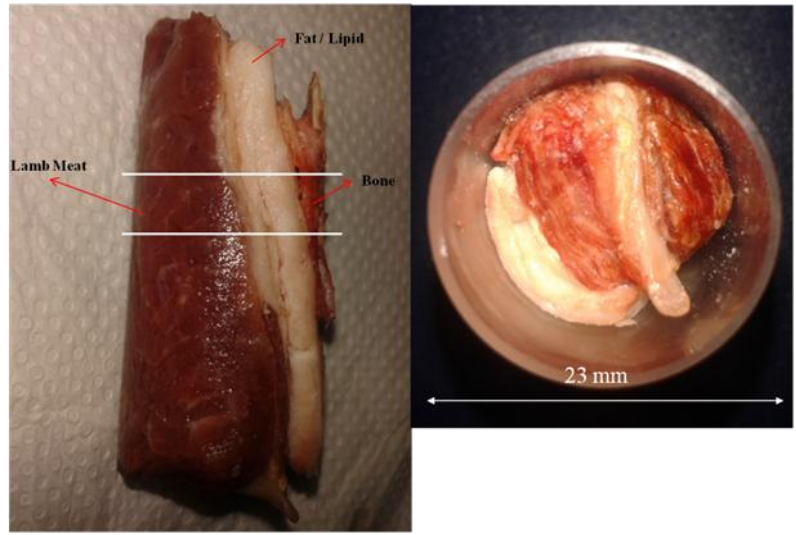

**Figure 2** *Left:* **The selected region from the meat has been scanned in the MARS system. The region included meat, fat and bone (referring calcium).** *Right:* **Setting of meat inside 23 mm phantom for scanning.**

## 3. Results

### 3.1 Post-acquisition Processing

The raw data in DICOM format from the scanner was processed using the MARS Image Processing Suite. The processing performs DICOM file conversion (DCM2mPPC) and pre-reconstruction processing (mPPC), which includes flat field and dark field corrections as shown in Figure 3, followed by ring filtration of the data.

Flat field is a standard method used for the correction of fixed pattern noise and therefore, it is necessary to average a large number of flat-field images to calculate the fixed-pattern correction image. Ring filtration is applied on corrected frames to reduce ring artefacts. Finally, the unstitched ring filtered frames were used to reconstruct the 3D volume in linear attenuation by using the MARS-Algebraic Reconstruction Technique (mART). CT reconstruction was performed in pseudo-spectral mode (15-50 keV, 20-50 keV etc). The raw, partially processed and fully processed data of lamb tissue along with the phantom calibration data can be found at [http://hdl.handle.net/10092/8531].

### 3.2 Hounsfield Units Calibration and Visualization

Information from the calibration phantom has been used to correlate between the lipid (fat), water (tissue) and calcium (bone) regions in the meat sample. Images of the sample showing the spectral response of transverse slice 58 with increasing threshold energy are shown in Figure 4. The CT number is quantified in spectral Hounsfield Unit by normalizing CT attenuation (commonly known as linear attenuation) to water and air to account for the differing attenuation at different energies [37]. A Hounsfield Unit is based on a scale of -1000 of air to +3000 of dense bone, with water being 0.

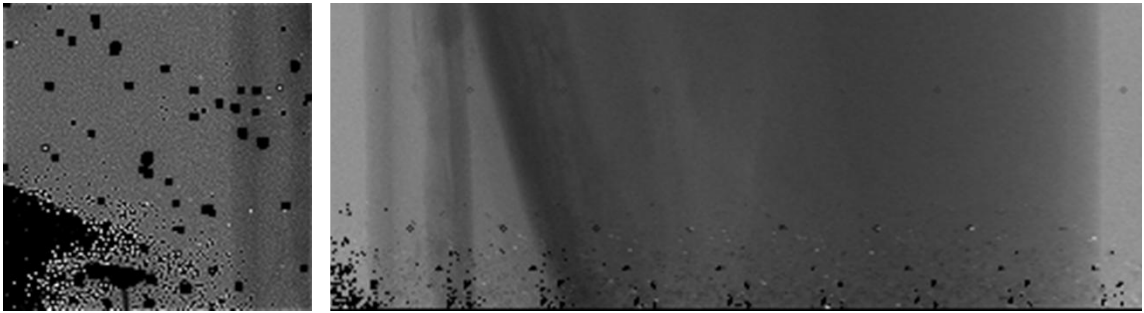

**Figure 3** *Left:* **A raw projection image for a single camera position. The dark pixels are non-responding pixels due to surface damage of the sensor layer, ASIC faults or bad bump bond connections. It is possible to interpolate the missing data, e.g. by linear interpolation across the non-responding pixels. However, this region is considered to be one of the main causes of undesirably longer scans.** *Right:* **The corresponding full field of view (FOV), which shows that oversampling from various camera positions and large number of projection images taken at different gantry angles, collects sufficient data to compensate for non-responding pixels in the image. The corrected image is the flat field image across the FOV without further data processing.**

The quantitative diagnostic information (decomposed volume) about the material composition of the sample has been extracted from MARS scans by applying material decomposition (MARS-MD) at high spatial resolution (<100 micron). MD techniques typically use either linear least-squares or maximum likelihood algorithms which compare the spectral signal to the unique spectral signatures of various materials of interest (such as lipid, water, calcium and any radio-pharmaceuticals which may have been used). In this study a constrained linear least squares algorithm is applied in the image domain using material attenuation information obtained from the calibration phantom containing the known materials described earlier. Figure 5 shows the classification of fat-like, water-like and calcium-like densities from randomly selected CT slice 100. A composite RGB image based on fat-like, water-like and calcium-like densities has been generated showing colourful discrimination of fat (blue) from soft tissue (red) and calcium (yellow) (see Figure 5). There is some cross talk observed in fat and soft tissue decomposed images which could be due to Compton Scattering. However, this is not considered to be a problem as the MARS-MD algorithm is currently in its development phase. It is expected that, in the near future, cross talk could be subtracted by comparing it with a high Z material image from soft tissue image. After material decomposition, 3D volume rendering of the whole scanned volume has been performed and shown in Figure 6.

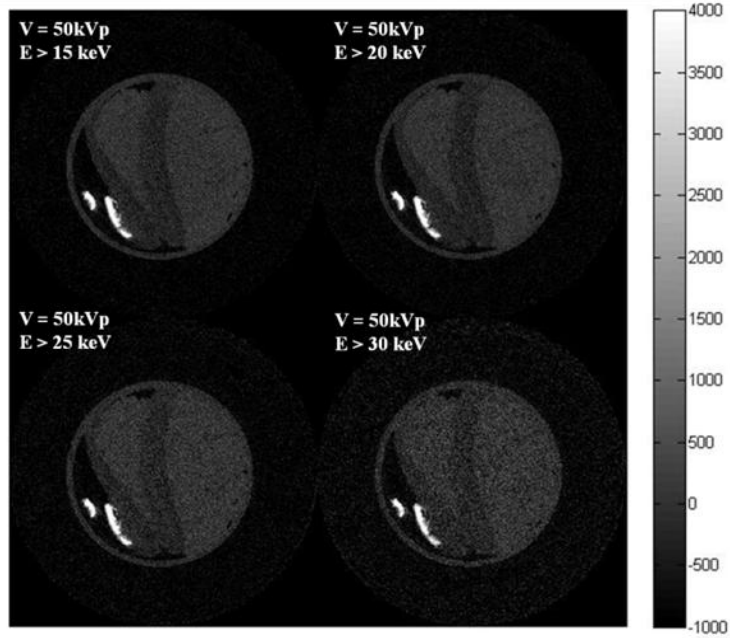

**Figure 4** Image showing transverse slice 58 with increasing threshold energies. The colour-map represents spectral HU ranging from -1000 to 4000.

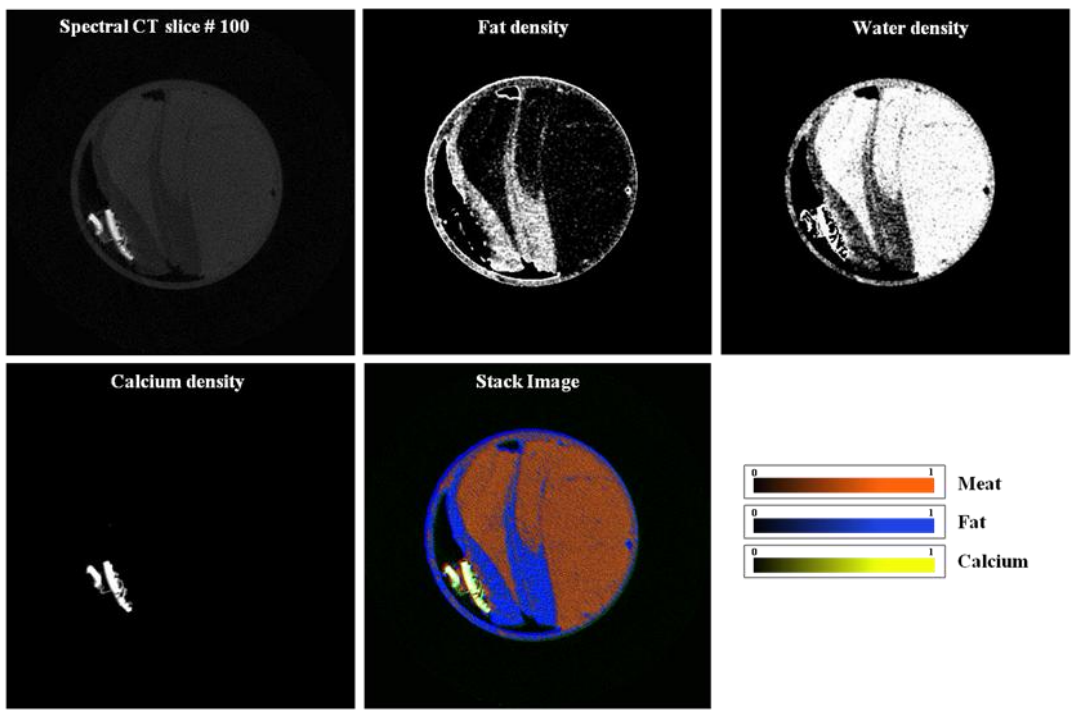

**Figure 5** Classification of spectral CT image into individual elements. Top *(left to right)*: spectral CT image of slice 100 with one of the energy bins, image with fat density and image with water density (referring soft tissue). Bottom *(left to right)*: image with calcium density and composite image of fat (blue) and water densities (red) and bone (yellow). Scale is relative to concentrations used to calibrate material decomposition.

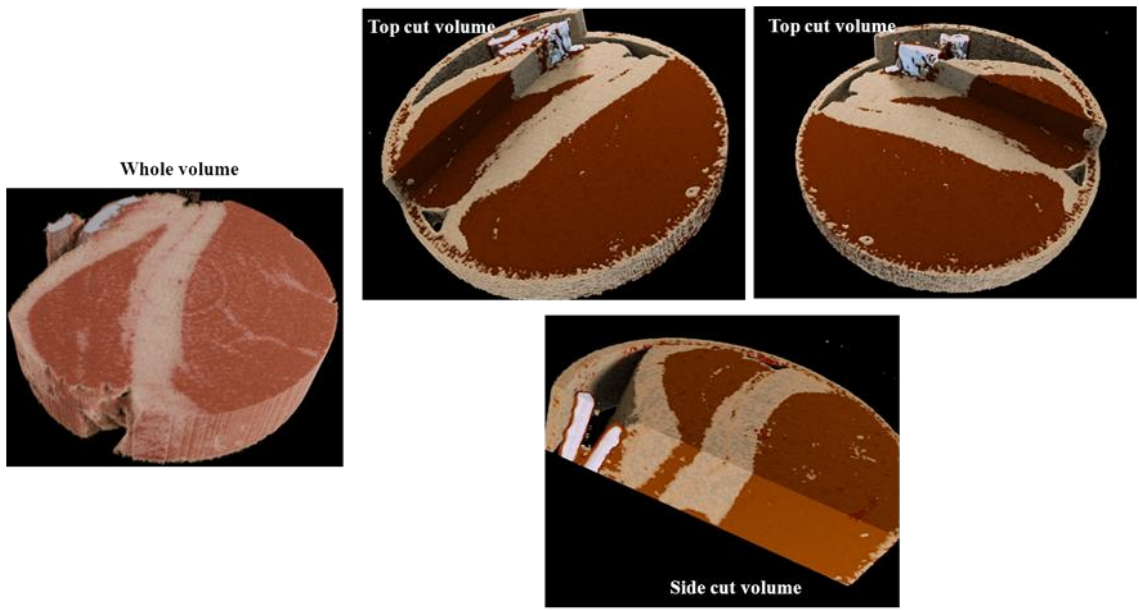

**Figure 6 3D volume rendering of lamb meat using the MARS system. A clear separation within meat structure between fat (off white colour scheme), meat (reddish), fat marbling and bone (showing calcium in white) can be seen.**

## 4. Discussion and Conclusion

In this experimental study, spectral imaging of tissue has been performed without using high Z contrast agents. We have shown that lipid-like, water-like and calcium-like components of tissue can be distinguished from each other at high spatial resolution (~100 micron). Dual-energy CT systems are limited to two energies and materials with closely related attenuation profiles, such as soft tissue (effective Z ≈ 7.5) and fat (effective Z ≈ 6.5), may be distinguished from each other but the methodology is difficult to perform [38]. However, the MARS scanner, incorporated with the Medipix3RX detector, divides a single wide spectrum into separate energy bins to extract energy information from the x-ray beam with fewer artefacts [33]. In a 3D volume rendering of the meat sample (see Figure 6), a clear distribution of tissue, fat and bone can be seen, including fat marbling in the muscle. These techniques are critical for preclinical research projects in the characterization of atherosclerotic plaque, fatty liver diseases (metabolic syndrome), breast imaging, and functional imaging with nanoparticles that are currently underway by various MARS researchers. High soft tissue sensitivity is also attractive for the meat industry by allowing more accurate portioning and identifying low and high fat regions.

A shortcoming of the study is the limited number of biological test materials examined. However, it is important to mention that firstly, this was not an extensive study of the contrast/noise aspects, but was done to show clinical radiology level CT images, produced with the MARS scanner and Medipix3RX camera, that clearly discriminated the main constituents of the body such as fat, bone, and soft tissues. And secondly, the intention was to share the raw, partially and fully processed data with others so they can process it using their familiar routines.

For more effective material decomposition, the energy bins used for material decomposition should ideally be narrow and separated [8, 39, 40]. However with an x-ray source and current detectors, when narrow bins are used, a large fraction of the detected x-ray counts is lost and statistical noise is increased. The selection of appropriate energy bins is an ongoing topic of research and is outside the scope of this work.

To explore full spectroscopic molecular imaging by MARS scanners, energy calibration of the Medipix detectors and their optimization need to be performed at the individual pixel level. While writing this, a detailed study is being done on individual pixel energy calibration by several of the MARS group members. CdTe versions of Medipix efficiently measure photons up to 140 keV and have similar detection efficiency to most other x-ray detectors used in CT. This energy range is also suitable for human imaging, enabling an eventual translation of this program's work into spectral molecular human imaging by using a larger Medipix3 detector array.

## Acknowledgments

This project was funded by Ministry of Business, Innovation and Employment (MBIE), New Zealand under contract number UOCX0805. The authors would like to thank all members of MARS-CT project, the Medipix2 collaboration, and the Medipix3 collaboration. In particular we acknowledge the CERN based Medipix3 designers Michael Campbell, Lukas Tlustos, Xavier Llopart, Rafael Ballabriga and Winnie Wong; and the Freiburg material scientists Michael Fiederle, Alex Fauler, Simon Procz, Elias Hamann and Martin Pichotka of Freiburger Materialforschungszentrum and X-ray Imaging Europe GmbH.